\documentclass[english,prl,twocolumn,amsfonts, tight]{revtex4}
\usepackage[T1]{fontenc}
\usepackage[latin1]{inputenc}
\usepackage{graphicx}
\usepackage{amssymb}

\makeatletter

\providecommand{\LyX}{L\kern-.1667em\lower.25em\hbox{Y}\kern-.125emX\@}

\providecommand{\tabularnewline}{\\}

\usepackage{babel}
\makeatother
\begin{document}

\preprint{This line only printed with preprint option}

\title{Graphene bilayer with a twist: electronic structure}

\author{J.~M.~B. Lopes dos Santos$^1$, N. M. R. Peres$^2$, and A. H. Castro Neto$^3$}


\affiliation{$^1$ CFP and Departamento de F\'{\i}sica, Faculdade de Ci\^{e}ncias,
Universidade do Porto, 4169-007 Porto, Portugal\\
$^2$ Centro de F\'{\i}sica and Departamento de F\'{\i}sica, Universidade
do Minho, P-4710-057, Braga, Portugal\\
$^3$ Department of Physics, Boston University, 590 Commonwealth Avenue,
Boston, MA 02215,USA}

\begin{abstract}
Electronic properties of bilayer and multilayer graphene have generally
been interpreted in terms of $AB$ or Bernal stacking. However, it
is known that many types of stacking defects can occur in natural
and synthetic graphite; rotation of the top layer is often seen in
scanning tunneling microscopy (STM) studies of graphite. In this paper we
consider a graphene bilayer with a relative small angle rotation between
the layers and calculate the electronic structure near zero energy
in a continuum approximation. Contrary to what happens in a $AB$
stacked bilayer and in accord with observations in epitaxial graphene we find:
(a) the low energy dispersion is linear, as in a single layer, but
the Fermi velocity can be significantly smaller than the single layer
value; (b) an external electric field, perpendicular to the layers,
does not open an electronic gap. 
\end{abstract}
\maketitle

\paragraph*{Introduction.}

Graphene is a two-dimensional (2D) carbon material, which takes the
form of a planar honeycomb lattice of $sp^{2}$ bonded carbon atoms.
It can be considered as a building block for other allotropes of carbon,
such as graphite, fullerenes, and carbon nanotubes and it was first 
isolated by micro-mechanical cleavage of graphite in 2004 \cite{NGM+04d_short, NJS+05_short}.
This method also produces samples composed of two (bilayer) or more
atomic layers of graphene (few layer graphene, FLG). FLG samples can
also be grown epitaxially by thermal decomposition of the surface of SiC 
\cite{BergerC._jp040650f_short}. 

Single layer (SLG) and bilayer (BLG) graphene are both gapless semi-metals,
if undoped, but whereas carriers in SLG have linear dispersion (leading
to Dirac cones in energy momentum space)\cite{NGM+05_short},
in BLG the dispersion is quadratic \cite{MF06_short}. The quantization rules
for the integer quantum Hall effect are different for SLG 
\cite{ZTS+05_short,PGN06b,NGM+05_short} and BLG \cite{NMM+06b_short}. 
A controllable gap can be opened with
an external electric field in BLG, a fact that makes it particularly
interesting for applications \cite{castro-2006_short,McC06}.

The properties of BLG have been interpreted under the assumption that
the stacking of the two layers takes the form of $AB$ or Bernal stacking,
the most common in graphite. Nevertheless, $AB$ stacking is
not the only form of stacking found in graphite. Naturally occurring
and synthetic (HOPG) crystals
usually present a variety of defects which affect stacking order in
the $c$ direction. Turbostratic graphite is modeled by stacking graphene
layers with random relative translations and rotations \cite{CML92};
rotation of the top layer with respect to the bulk is quite common
in the surface of graphite and results in the formation of superlattices
clearly seen in STM images as Moire patterns \cite{PD05_short,RK93}. Recent
detailed structural studies of epitaxially grown FLG \cite{hass-2007_short}
rule out $AB$ stacking and reveal the presence of significant orientational
disorder of the graphene with respect to the underlying SiC substrate \cite{HFL+06_short}.
The influence of the type of stacking on the electronic structure
in multilayer graphene has been stressed by Guinea {\it et al.}
\cite{GNP06}. 

In this work we discuss the electronic structure of a bilayer with
a relative, small-angle, rotation of the two graphene planes. We derive
angles for the formation of  periodic Moire superlattices and
formulate a continuum electronic description in terms of massless
Dirac fermions, coupled by a slowly varying periodic inter-layer hopping.
We find a low energy electronic structure quite different from that
of $AB$ stacked bilayer, with massless Dirac fermions, but with a 
Fermi velocity ($v_{F}$)
substantially reduced with respect to SLG. Moreover, we show that 
an external electric field
does not open a gap in the electronic spectrum. 
These results are all in accord with observations
in epitaxially grown graphene, which reveal much the same electronic behavior
as SLG in angle resolved photoemission (ARPES) \cite{HFL+06_short,bostwick-2006_short,ZGG+06_short},
transport \cite{BSL+06_short}, and infrared spectroscopy \cite{SMP+06b_short}
and display systematically reduced values of $v_{F}$ relative to SLG \cite{deheer07}.

\paragraph*{Geometry.}

The two sublattices in layer 1 are denoted by $A$ and $B$ and in
layer 2 by $A'$ and $B'.$ In an $AB$ stacked bilayer $A$ and $B'$
atoms have the same horizontal positions, $i\mathbf{a}_{1}+j\mathbf{a}_{2}$
($i,j$ integers), where $\mathbf{a}_{1}=(1/2,\sqrt{3}/2)a_{0}$,
$\mathbf{a}_{2}=(-1/2,\sqrt{3}/2)a_{0}$ are the Bravais lattices
basis vectors and $a_{0}$ ($\approx 2.46\,\textrm{\AA}$) is the lattice constant.
The SLG Dirac points are located at $\mathbf{K}=-\mathbf{K}'=(4\pi/3,0)/a_{0}$.
The vertical displacement between the layers is $\mathbf{c}_{0}$
($\approx 3.35\,\textrm{\AA}$). 

For simplicity we consider rotations of layer 2 about a site occupied
by a $B'$ atom (directly opposite an $A$ atom, in the $c$ direction):
a commensurate structure is obtained if a $B'$ atom is moved by the
rotation to a position formerly occupied by an atom of the same kind.
The Moire pattern is periodic and the translation from the origin
(center of rotation) to the $B'$ atom's current position is a symmetry
translation. From this we can derive
 a condition  for the angle $\theta_i$  of  a commensurate
rotation: 
\begin{equation}
\cos(\theta_{i})=\frac{3i^{2}+3i+1/2}{3i^{2}+3i+1},\qquad i=0,1,2\dots.\label{eq:cos_theta_i}\end{equation}
The superlattice basis vectors are: 
\begin{eqnarray}
\mathbf{t}_{1} & = & i\mathbf{a}_{1}+(i+1)\mathbf{a}_{2} \, ,
\nonumber \\
\mathbf{t}_{2} & = & -(i+1)\mathbf{a}_{1}+(2i+1)\mathbf{a}_{2} \, ,
\label{eq:superlatice_basis}\end{eqnarray}
($i=0$ is an $AA$ stacked bilayer). The lattice constant of the
superlattice is $L=\left|\mathbf{t}_{1}\right|=\sqrt{3i^{2}+3i+1} \, a_0$. 
STM measurements of the surface of graphite \cite{RK93} 
observed superlatices with periods of
$L=66\,\textrm{\AA}$ and angles $\theta=2.1^{\mathrm{o}}$ corresponding
to $i=15$ in (\ref{eq:cos_theta_i}) and a unit cell with 2884 atoms, 
making {\it ab initio} descriptions rather impractical. 
The reciprocal lattice vectors are:
\begin{eqnarray}
\mathbf{G}_{1} & = & \frac{4\pi}{3\left(3i^{2}+3i+1\right)}\left[\left(3i+1\right)\mathbf{a}_{1}+\mathbf{a}_{2}\right] \, ,
\label{eq:g1_superlattice}\\
\mathbf{G}_{2} & = & \!\! \frac{4\pi}{3\left(3i^{2}+3i+1\right)}\left[-\left(3i+2\right)\mathbf{a}_{1}+\left(3i+1\right)\mathbf{a}_{2}\right].
\label{eq:g2_superlattice}
\end{eqnarray}

\paragraph*{Continuum description.}

The Hamiltonian for the bilayer with a twist has the form $\mathcal{H}_{1}+\mathcal{H}_{2}+\mathcal{H}_{\perp}$,
with the intra-layer Hamiltonian, $\mathcal{H}_{1}+\mathcal{H}_{2}$, given by (we use units such that $\hbar=1$):
\begin{eqnarray}
\mathcal{H}_{1} & = & -t\sum_{i}c_{A}^{\dagger}(\mathbf{r}_{i})\left[c_{B}(\mathbf{r}_{i}+\mathbf{s}_{0})+c_{B}(\mathbf{r}_{i}+\mathbf{s}_{0}-\mathbf{a}_{1})\right.
\nonumber 
\\
 & + & \left.c_{B}(\mathbf{r}_{i}+\mathbf{s}_{0}-\mathbf{a}_{2})\right]+ {\rm h.c.} \, ,
\label{eq:ham1}
\\
\mathcal{H}_{2} & = & -t\sum_{j}c_{B'}^{\dagger}(\mathbf{r}_{j})\left[c_{A'}(\mathbf{r}_{j}-\mathbf{s}_{0}')+c_{A'}(\mathbf{r}_{j}-\mathbf{s}_{0}'+\mathbf{a}_{1}')\right.\nonumber \\
 & + & \left.c_{A'}(\mathbf{r}_{j}-\mathbf{s}_{0}'+\mathbf{a}_{2}')\right]+ {\rm h.c.},
\label{eq:ham2}\end{eqnarray}
where $c_{\alpha}(\mathbf{r})$ is the
destruction operator for the state in sublattice $\alpha$ at horizontal
position $\mathbf{r}$; $\alpha=A,B$ in layer 1 and $\alpha=A',B'$
in layer 2; $\mathbf{a}_{1}'$ and $\mathbf{a_{2}}'$ are obtained
from $\mathbf{a}_{1}$ and $\mathbf{a}_{2}$ by a rotation by $\theta$
about the origin; $\mathbf{r}_{i}=m\mathbf{a}_{1}+n\mathbf{a}_{2}$
for $\mathcal{H}_{1}$ and $\mathbf{r}_{j}=r\mathbf{a}_{1}'+s\mathbf{a}_{2}'$
for $\mathcal{H}_{2}$ ($m,n,r,s,$ integers); $\mathbf{s}_{0}=(\mathbf{a}_{1}+\mathbf{a}_{2})/3$
and $\mathbf{s}'_{0}=(\mathbf{a}'_{1}+\mathbf{a}'_{2})/3$. 

To study the low energy spectrum near the $\mathbf{K}$ ($\mathbf{K'}$) point,
we go to the continuum limit, with the standard replacement $c_{\alpha}(\mathbf{r})\to v_{c}^{1/2}\psi_{1,\alpha}(\mathbf{r})\exp(i\mathbf{K}\cdot\mathbf{r})$
where $\psi_{1,\alpha}(\mathbf{r})$ is a slowly varying field on
scale of the lattice constant ($v_{c}$ is the unit cell
volume). Due to the rotation, the wave vector in layer 2 is
shifted to $\mathbf{K}^{\theta}=4\pi(\cos\theta,\sin\theta)/(3a_{0})$,
so $c_{\alpha'}(\mathbf{r})\to v_{c}^{1/2}\psi_{2,\alpha}(\mathbf{r})\exp(i\mathbf{K}^{\theta}\cdot\mathbf{r}).$
For small angles of rotation the modulation of inter-layer hopping
has a long wavelength and the coupling between different valleys ($\mathbf{K}$
and $\mathbf{K}'$) can be ignored. Hence, in the long-wavelength limit the decoupled
Hamiltonian can be written as:
\begin{eqnarray}
\mathcal{H}_{1} & = & v_{F}\sum_{k}\psi_{1,\mathbf{k}}^{\dagger}\mathbf{\tau\cdot k}\psi_{1,\mathbf{k}} \, ,
\label{eq:ham1_cont}
\\
\mathcal{H}_{2} & = & v_{F}\sum_{k}\psi_{2,\mathbf{k}}{}^{\dagger}\mathbf{\tau^{\theta}\cdot}\mathbf{k}\psi_{2,\mathbf{k}},
\label{eq:ham2_cont}\end{eqnarray}
where $v_{F}=at\sqrt{3}/2$ and $\tau = (\tau_x,\tau_y)$ are Pauli matrices.
The coordinate axes have been chosen to coincide with those of layer
1, so the Hamiltonian of layer 2 involves a extra rotation by $\theta$,
the angle between the two layers and $\tau^{\theta}=e^{+i\theta\tau_{z}/2}(\tau_{x},\tau_{y})e^{-i\theta\tau_{z}/2}$. 

To model the inter-layer coupling, $\mathcal{H}_{\perp}$, we retain
hopping from each site in layer 1 to the closest sites of layer 2
in either sub-lattice. We denote by $\mathbf{\delta}^{\beta'\alpha}(\mathbf{r})$
the horizontal (in-plane) displacement from an atom of layer 1, sub-lattice
$\alpha$($\alpha=A,B)$ and position $\mathbf{r}$, to the closest
atom in layer 2, sub-lattice $\beta'$ ($\beta'=A',B'$). Denoting
by $t_{\perp}\left(\mathbf{\delta}\right)$ the hopping between $p_{z}$
orbitals with a relative displacement $\mathbf{c}_{0}+\mathbf{\delta}$,
one gets\begin{equation}
\mathcal{H}_{\perp}=\sum_{i,\alpha,\beta'}t_{\perp}\left(\mathbf{\delta}^{\beta'\alpha}(\mathbf{r}_{i})\right)c_{\alpha}^{\dagger}(\mathbf{r}_{i})c_{\beta'}\left(\mathbf{r}_{i}+\delta^{\beta'\alpha}(\mathbf{r}_{i})\right)+ {\rm h.c.}\label{eq:h_perp1}\end{equation}
where $t_{\perp}\left(\mathbf{\delta^{\alpha\beta}}(\mathbf{r})\right)\equiv t_{\perp}^{\alpha\beta}(\mathbf{r})$,
is the inter-layer, position dependent, hopping between $p_{z}$ orbitals
with a relative displacement $\mathbf{c}_{0}+\mathbf{\delta}$; $\Delta\mathbf{K}=\mathbf{K}^{\theta}-\mathbf{K}$
is the relative shift between corresponding Dirac wavectors in the
two layers; $\phi_{i,k,\alpha}=\psi_{i,k\pm\Delta K/2,\alpha}$ is
the Fourier component of $\psi_{i,\alpha}\left(\mathbf{r}\right)$
for momentum \emph{}$\mathbf{k}\pm\Delta\mathbf{K}/2$, the plus sign
applying in layer 1 and the minus one in layer 2. With this choice,
the Dirac fields $\phi_{i,k,\alpha}$ with the same $\mathbf{k}$
vector in both layers correspond to the same plane waves in the original
lattice; the Dirac cones occur at $\mathbf{k}=-\Delta\mathbf{K}/2$
in layer 1 and $\Delta\mathbf{K}/2$ in layer 2. Replacing the operators
in eq.~(\ref{eq:h_perp1}) with the Dirac fields, using $\psi_{i,\beta}(\mathbf{r}+\mathbf{\delta}^{\beta\alpha}(\mathbf{r}))\approx\psi_{i,\beta}(\mathbf{r})$,
since the Dirac fields are slowly varying on the lattice scale, and
Fourier transforming, the low energy effective Hamiltonian, near $\mathbf{K}$,
is\begin{eqnarray}
\mathcal{H} & = & v_{F}\sum_{k,\alpha\beta}\phi_{1,k,\alpha}^{\dagger}\mathbf{\tau}_{\alpha\beta}\cdot\left(\mathbf{k}+\frac{\Delta\mathbf{K}}{2}\right)\phi_{1,k,\beta}\nonumber \\
 & + & v_{F}\sum_{k,\alpha,\beta}\phi_{2,k,\alpha}^{\dagger}\mathbf{\tau}_{\alpha\beta}^{\theta}\cdot\left(\mathbf{k}-\frac{\Delta\mathbf{K}}{2}\right)\phi_{2,k,\beta}\nonumber \\
 & + & \left(\sum_{\alpha,\beta}\sum_{\mathbf{k},\mathbf{G}}\tilde{t}_{\perp}^{\beta\alpha}(\mathbf{G})\phi_{1,k+G,\alpha}^{\dagger}\phi_{2,k,\beta}+ {\rm h.c.}\right).
\label{eq:complete_ham2}\end{eqnarray}
For commensurate structures, the function $t_{\perp}^{\alpha\beta}(\mathbf{r})\exp\left(i\mathbf{K}^{\theta}\cdot\delta^{\alpha\beta}(r)\right)$
is periodic and has nonzero Fourier components only at the vectors
$\mathbf{G}$ of the reciprocal lattice :
\begin{equation}
\tilde{t}_{\perp}^{\alpha\beta}(\mathbf{G})=\frac{1}{V_{c}}\int_{v_c}d^{2}r\, t_{\perp}^{\alpha\beta}(\mathbf{r})e^{i\mathbf{K}^{\theta}\cdot\delta^{\alpha\beta}(r)}e^{-i\mathbf{G}\cdot\mathbf{r}}.\label{eq:t_perp_G}\end{equation}
The integral is over the unit cell of the superlattice
and it will ultimately be calculated by a sum over the sites
of the Wigner-Seitz unit cell since $t_{\perp}^{\alpha\beta}(\mathbf{r})$ is
only defined at those points. This Hamiltonian describes two sets of
relativistic Dirac fermions (with shifted degeneracy points) coupled
by a periodic perturbation.

\paragraph*{Fourier amplitudes of inter-layer coupling.}
To determine the hopping $t_{\perp}(\delta)$ as a function of the
horizontal shift $\mathbf{\delta}^{\alpha\beta}(\mathbf{r})$ we express
it in the Slater-Koster parameters, $V_{pp\sigma}(d)$ and $V_{pp\pi}(d)$,
where $d$ is the distance between the two atomic centers, $d=\sqrt{c_{0}^{2}+\delta^{2}}$.
For the $d$ dependence of $V_{pp\sigma}(d)$ and $V_{pp\pi}(d)$
we used the parametrization of ref.~\cite{PhysRevB.53.979-b_short}.
$V_{pp\pi}(a_{0}/\sqrt{3})$, is the in-plane nearest neighbor hopping,
$t$, and $V_{pp\sigma}(c_{0})$ if the inter-layer hopping, $t_{\perp}$,
in an $AB$ stacked bilayer. The contribution of $V_{pp\pi}$ turns
out to be negligible and $t_{\perp}(\delta)$ is proportional
to $t_{\perp}$: for $\delta=a_{0}/\sqrt{3}$, $t_{\perp}(\delta)/t_{\perp}\approx0.4$.
We have calculated $\delta^{\alpha\beta}(\mathbf{r})$ numerically
for any angle of rotation. Using various symmetries and relations valid in the
limit $a_{0}\ll L$ (small angles) we were able to derive the results
of Table~\ref{tab:fourier_amplitudes}. The values of $\tilde{t}_{\perp}^{BA}(\mathbf{G})$
are equal and real, by symmetry, for $\mathbf{G}=0$, $\mathbf{G}=-\mathbf{G}_{1}$
and $\mathbf{G}=-\mathbf{G}_{1}-\mathbf{G}_{2}$ and much smaller
for all other $\mathbf{G}$ vectors. The remaining Fourier amplitudes
can be expressed in terms of $\tilde{t}_{\perp}^{BA}(\mathbf{G})$.%
\begin{figure}
\begin{centering}
\includegraphics[width=0.5\columnwidth]{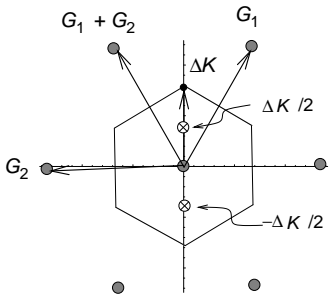}
\par\end{centering}
\caption{First Brillouin zone (FBZ) of the super-lattice centered at mid-point between
Dirac points $\mathbf{K}$ and $\mathbf{K}^{\theta}$. Note that the
zero energy states of the two layers, $\mathbf{k}=-\Delta\mathbf{K}/2$
and $\mathbf{k}=\Delta\mathbf{K}/2$, marked with $\otimes$, are
half-way to the zone boundary; $\Delta\mathbf{K}$ is a vertex of
the FBZ. \label{fig:First-Brillouin-zone}}
\end{figure}
\begin{table}
\begin{centering}
\begin{tabular}{|c||c|c|c|}
\hline 
$\mathbf{G}$&
\textbf{\Large \hspace{.5cm}\rule[-6pt]{0pt}{18pt}}0\textbf{\Large \hspace*{.5cm}}&
$\mathbf{-G}_{1}$&
$-\mathbf{G}_{1}-\mathbf{G}_{2}$\tabularnewline
\hline
\hline 
$\tilde{t}_{\perp}^{BA}(\mathbf{G})$&
\textbf{\Large \rule[-6pt]{0pt}{18pt}}$\tilde{t}_{\perp}$&
$\tilde{t}_{\perp}$&
$\tilde{t}_{\perp}$\tabularnewline
\hline 
$\tilde{t}_{\perp}^{AB}(\mathbf{G})$&
\textbf{\Large \rule[-6pt]{0pt}{18pt}}$\tilde{t}_{\perp}$&
$e^{-i2\pi/3}\tilde{t}_{\perp}$&
$e^{i2\pi/3}\tilde{t}_{\perp}$\tabularnewline
\hline 
$\tilde{t}_{\perp}^{AA}(\mathbf{G})$&
\textbf{\Large \rule[-6pt]{0pt}{18pt}}$\tilde{t}_{\perp}$&
$e^{i2\pi/3}\tilde{t}_{\perp}$&
${e^{-i2\pi/3}\tilde{t}}_{\perp}$\tabularnewline
\hline 
$\tilde{t}_{\perp}^{BB}(\mathbf{G})$&
\textbf{\Large \rule[-6pt]{0pt}{18pt}}$\tilde{t}_{\perp}$&
${e^{i2\pi/3}\tilde{t}}_{\perp}$&
$e^{-i2\pi/3}\tilde{t}_{\perp}$\tabularnewline
\hline
\end{tabular}
\par\end{centering}

\caption{The most important Fourier amplitudes are shown in this table (all
others are smaller by at least a factor of 5 for angles smaller than
10º). The first and second lines express exact results: $\tilde{t}_{\perp}$
is real. In the last two lines there are corrections to these results
of order $a_{0}/L$ where $L$ is the period of the superlattice.
\label{tab:fourier_amplitudes}}
\end{table}

\paragraph*{Results and discussion.}

\begin{figure}
\begin{centering}
\includegraphics[width=1\columnwidth]{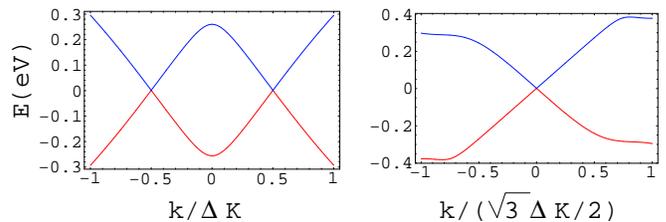}
\par\end{centering}

\caption{The energy $\epsilon_{k}$ of the two states with smaller $\left|\epsilon_{k}\right|$
for $\theta=3.9^{0}$ ($i=8$); panel (a): $\mathbf{k}$ varying form
$-\Delta\mathbf{K}$ to $\Delta\mathbf{K}$ (two vertexes of the FBZ)
along the line passing the degeneracy points, $-\Delta\mathbf{K}/2$
to $\Delta\mathbf{K}/2$; panel (b): along a line parallel to $\mathbf{G}_{2}$
passing $\Delta\mathbf{K}/2$.\label{fig:disp_relation} }
\end{figure}

In the absence of the inter-layer coupling, $\mathcal{H}_{\perp}$,
states with energy close to zero occur at $\mathbf{k}=-\Delta\mathbf{K}/2$
in layer 1 and $\mathbf{k}=+\Delta\mathbf{K}/2$ in layer 2. The results
of Table~\ref{tab:fourier_amplitudes} imply that the states of momentum
$\mathbf{k}$ in layer 1 are coupled directly only to states of layer
2 of momentum $\mathbf{k}$, $\mathbf{k}+\mathbf{G}_{1}$ and $\mathbf{k}+\mathbf{G}_{1}+\mathbf{G}_{2}$;
conversely the states of momentum $\mathbf{k}$ in layer 2 only couple
to states $\mathbf{k}$, $\mathbf{k}-\mathbf{G}_{1}$ and $\mathbf{k}-\mathbf{G}_{1}-\mathbf{G}_{2}$.
To investigate the spectrum at a momentum $\mathbf{k}$ close to zero
energy, we truncated the Hamiltonian to include only these six momentum
values (three for each layer) giving a $12\times12$ matrix to diagonalize.
The geometry of the first Brillouin zone (FBZ) of the superlattice (fig.~\ref{fig:First-Brillouin-zone})
implies that the states near the degeneracy point in either layer couple
only to states of energies $\pm v_{F}\Delta K=\pm v_{F}K\times2\sin(\theta/2)$
where $\Delta K=\left|\Delta\mathbf{K}\right|$ and $K=4\pi/(3a_{0})$.
This turns out to be the essential difference between this problem
and that of the unrotated bilayer. In the latter, the degeneracy points
of both layers occur at the same momentum and the inter-layer hopping
couples two doublets of zero energy states. In the present case we
have one doublet of zero energy states coupling to three pairs of
states at finite energies, $\pm v_{F}\Delta K$. As a result,
the linear dispersion near zero energy is retained. In Fig.~\ref{fig:disp_relation}
we plot the energies of the states with smallest $\left|\epsilon_{k}\right|$
along two lines in the FBZ; the parameters are $t_{\perp}=0.27\,\mathtt{eV}$
\cite{castro-2006_short} and $\theta=3.9^{0}$ ($i=8,$ $L=36\,\textrm{\AA})$,
which give $v_{F}\Delta K\approx0.76\,\mathtt{eV}$ and $\tilde{t}_{\perp}=0.11\,\mathtt{eV}$.

The persistence of the Dirac cones can be understood by considering
the limit where $\tilde{t}_{\perp}/(v_{F}\Delta K)\ll1$ (in
the situation represented in fig.~\ref{fig:disp_relation}, $\tilde{t}_{\perp}/(v_{F}\Delta K)\approx0.14$).
Consider, for instance, the vicinity of the degeneracy point of layer
1, $\mathbf{k}=-\Delta\mathbf{K}/2+\mathbf{q}$. It is clear that
the Hamiltonian $\mathrm{H}(\mathbf{k})$ has the form $\mathrm{H}(\mathbf{k})=\mathbf{\mathrm{H}}(-\mathbf{K}/2)+\mathrm{V(\mathbf{q})}$
with $\mathrm{V}(\mathbf{q})$ linear in $\mathbf{q}$. In $\mathbf{\mathrm{H}}(-\mathbf{K/2})$,
which contains the inter-layer coupling, the doublet at zero energy
couples with an amplitude $\sim\tilde{t}_{\perp}$ to six states (of
layer 2) with energies $\pm v_{F}\Delta K$. Using perturbation
theory one can derive an effective Hamiltonian in the space of the
zero energy doublet by considering the mixing of these six states
in layer 2 to first order in $\tilde{t}_{\perp}/\left(v_{F}\Delta K\right)$.
The degeneracy is not lifted, although there is a small shift in energy,
$\epsilon_{0}=6\tilde{t}_{\perp}^{2}\sin(\theta/2)/(v_{F}\Delta K)$.
For small $\mathbf{q}$ we can treat $\mathrm{V(\mathbf{q})}$ as
a perturbation in the subspace of this doublet: the effective Hamiltonian
matrix has the form characteristic of a Dirac cone\[
\mathrm{H_{eff}}=\left[\begin{array}{cc}
\epsilon_{0} & \tilde{v}_{F}q^{*}\\
\tilde{v}_{F}q & \epsilon_{0}\end{array}\right]\]
with $q=q_{x}+iq_{y}$. To second order in $\tilde{t}_{\perp}/v_{F}\Delta K$,
the renormalized Fermi velocity is given by $\tilde{v}_{F}/v_{F}=1-9\left(\tilde{t}_{\perp}/(v_{F}\Delta K)\right)^{2}$
. This significant depression of the value of the Fermi velocity $\tilde{v}_{F}$
relative to the value of single layer graphene is a tell-tale sign
of the presence of a bilayer with a twist. The perturbative results
slightly overestimates the downward renormalization of $v_{F}$, because
of the contributions of higher order terms in $\tilde{t}_{\perp}/\left(v_{F}\Delta K\right)$,
especially at smaller angles (smaller $\Delta K$, larger $\tilde{t}_{\perp}/\left(v_{F}\Delta K\right)$).
Ref. \cite{deheer07} reports several observations of values of
$v_{F}$ in the range $0.7\sim0.8\times10^{6}\,\mathtt{m}\,\mathtt{s}^{-1}$,
in epitaxial graphene, 20 to 30\% lower than in single layers. 

\begin{figure}
\begin{centering}
\includegraphics[width=0.5\columnwidth]{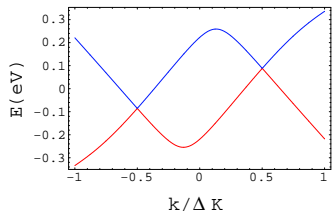}
\par\end{centering}

\caption{The energy $\epsilon_{k}$ of the two sates with smaller $\left|\epsilon_{k}\right|$
in the presence of a potential difference $V=0.3\,\mathtt{V}$ between
layers; $\mathbf{k}$ varies from $-\Delta\mathbf{K}$ to $\Delta\mathbf{K}$
(two vertexes of the FBZ) along the line passing the degeneracy points,
$-\Delta\mathbf{K}/2$ to $\Delta\mathbf{K}/2$; the remaining parameters
are the ones used in Fig.~\ref{fig:disp_relation}.\label{fig:disp_rel_field} }
\end{figure}

Another important consequence of the rotation between layers occurs
when there is an electric potential difference between layers. To
the Hamiltonian (\ref{eq:complete_ham2}) this adds a
term $\mathcal{V}_{ext}=-(V/2)\sum_{k,\alpha}\phi_{1,k,\alpha}^{\dagger}\phi_{1,k,\alpha}+(V/2)\sum_{k,\alpha}\phi_{2,k,\alpha}^{\dagger}\phi_{2,k,\alpha}$.
It is known that in the $AB$ stacked bilayer a gap opens in the spectrum
in the presence of an external electric field between layers \cite{castro-2006_short,McC06}.
However, it is clear from the discussion above that the cones present
in the bilayer with a twist are essentially the Dirac cones of each
layer perturbed by the admixture of states of the opposing layer,
which are distant in energy. As such, we expect that a potential difference
between the layers, $V$, should merely give rise to a relative shift
of the energies of the degeneracy points in each cone, at least as
long as $V<v_{F}\Delta K$. This expectation is borne by the
results shown in Fig.~\ref{fig:disp_rel_field}; the Dirac cones
are shifted but there is no gap in the spectrum. %

The results of Table~\ref{tab:fourier_amplitudes} imply that a small
angle rotation destroys the particle-hole symmetry of an $AB$ stacked
bilayer (with hopping only between $A$ and $B'$ atoms). The Fermi
level of an undoped sample need no longer be at zero energy; in turbostratic
graphite, for instance, it is shifted to $0.11\,\mathtt{eV}$ \cite{CML92}.
This calculation, being limited to energies close to zero, cannot
determine the absolute position of the Fermi Level as a function of
carrier concentration.

In conclusion, we presented a detailed geometrical description of
a bilayer with a relative rotation between the layers. We developed
a continuum description valid for small angles of rotation and analyzed
the energy spectrum close to zero energy. We found that the Dirac
cones of a single layer graphene remain present in the bilayer, but
with a significant reduction of the Fermi velocity especially for
very small angles of rotation. A new energy scale is introduced $v_{F}\Delta K=v_{F}K\times2\sin(\theta/2)$
where $K=4\pi/(3a_{0})$ and $\theta$ is the angle of rotation; the
dispersion relation is only linear for energies such that $\left|\epsilon_{k}\right|<v_{F}\Delta K$.
Unlike the case of the $AB$ stacked bilayer, a potential difference
between layers does not open a gap in the spectrum. These results
show that a small stacking defect such as a rotation can have a profound
effect on the low energy properties of the bilayer and are in accord
with several observations in epitaxial graphene. 

\begin{acknowledgments}
The authors would like to thank very useful discussions with C. Berger,
 E. H. Conrad, A. Geim, P. Guinea, J. Hass, W. de Heer, A. Lanzara and V.M. Pereira. JMBLS and NMRP acknowledge financial support from POCI 2010 via project
PTDC/FIS/64404/2006. A.H.C.N. was supported through NSF grant DMR-0343790.
\end{acknowledgments}

\begin{thebibliography}{22}
\expandafter\ifx\csname natexlab\endcsname\relax\def\natexlab#1{#1}\fi
\expandafter\ifx\csname bibnamefont\endcsname\relax
  \def\bibnamefont#1{#1}\fi
\expandafter\ifx\csname bibfnamefont\endcsname\relax
  \def\bibfnamefont#1{#1}\fi
\expandafter\ifx\csname citenamefont\endcsname\relax
  \def\citenamefont#1{#1}\fi
\expandafter\ifx\csname url\endcsname\relax
  \def\url#1{\texttt{#1}}\fi
\expandafter\ifx\csname urlprefix\endcsname\relax\def\urlprefix{URL }\fi
\providecommand{\bibinfo}[2]{#2}
\providecommand{\eprint}[2][]{\url{#2}}

\bibitem[{\citenamefont{{Novoselov {\it et al.}}}(2004)}]{NGM+04d_short}
\bibinfo{author}{\bibfnamefont{K.~S.} \bibnamefont{{Novoselov {\it et al.}}}},
  \bibinfo{journal}{Science} \textbf{\bibinfo{volume}{306}},
  \bibinfo{pages}{666} (\bibinfo{year}{2004}).

\bibitem[{\citenamefont{{Novoselov {\it et
  al.}}}(2005{\natexlab{a}})}]{NJS+05_short}
\bibinfo{author}{\bibfnamefont{K.~S.} \bibnamefont{{Novoselov {\it et al.}}}},
  \bibinfo{journal}{PNAS} \textbf{\bibinfo{volume}{102}},
  \bibinfo{pages}{10451} (\bibinfo{year}{2005}{\natexlab{a}}).

\bibitem[{\citenamefont{{Berger {\it et
  al.}}}(2004)}]{BergerC._jp040650f_short}
\bibinfo{author}{\bibfnamefont{C.}~\bibnamefont{{Berger {\it et al.}}}},
  \bibinfo{journal}{J. Phys. Chem. B} \textbf{\bibinfo{volume}{108}},
  \bibinfo{pages}{19912} (\bibinfo{year}{2004}).

\bibitem[{\citenamefont{{Novoselov {\it et
  al.}}}(2005{\natexlab{b}})}]{NGM+05_short}
\bibinfo{author}{\bibfnamefont{K.~S.} \bibnamefont{{Novoselov {\it et al.}}}},
  \bibinfo{journal}{Nature} \textbf{\bibinfo{volume}{438}},
  \bibinfo{pages}{197} (\bibinfo{year}{2005}{\natexlab{b}}).

\bibitem[{\citenamefont{McCann and Fal'ko}(2006)}]{MF06_short}
\bibinfo{author}{\bibfnamefont{E.}~\bibnamefont{McCann}} \bibnamefont{and}
  \bibinfo{author}{\bibfnamefont{V.~I.} \bibnamefont{Fal'ko}},
  \bibinfo{journal}{Phys. Rev. Lett.} \textbf{\bibinfo{volume}{96}},
  \bibinfo{pages}{086805} (\bibinfo{year}{2006}).

\bibitem[{\citenamefont{{Zhang {\it et al.}}}(2005)}]{ZTS+05_short}
\bibinfo{author}{\bibfnamefont{Y.~B.} \bibnamefont{{Zhang {\it et al.}}}},
  \bibinfo{journal}{Nature} \textbf{\bibinfo{volume}{438}},
  \bibinfo{pages}{201} (\bibinfo{year}{2005}).

\bibitem[{\citenamefont{Peres et~al.}(2006)\citenamefont{Peres, Guinea, and
  {Castro Neto}}}]{PGN06b}
\bibinfo{author}{\bibfnamefont{N.~M.~R.} \bibnamefont{Peres}},
  \bibinfo{author}{\bibfnamefont{F.}~\bibnamefont{Guinea}}, \bibnamefont{and}
  \bibinfo{author}{\bibfnamefont{A.~H.} \bibnamefont{{Castro Neto}}},
  \bibinfo{journal}{Phys. Rev. B} \textbf{\bibinfo{volume}{73}},
  \bibinfo{pages}{125411} (\bibinfo{year}{2006}).

\bibitem[{\citenamefont{{Novoselov {\it et al.}}}(2006)}]{NMM+06b_short}
\bibinfo{author}{\bibfnamefont{K.~S.} \bibnamefont{{Novoselov {\it et al.}}}},
  \bibinfo{journal}{Nature Physics} \textbf{\bibinfo{volume}{2}},
  \bibinfo{pages}{177} (\bibinfo{year}{2006}).

\bibitem[{\citenamefont{{Castro {\it et al.}}}(2006)}]{castro-2006_short}
\bibinfo{author}{\bibfnamefont{E.~V.} \bibnamefont{{Castro {\it et al.}}}}
  (\bibinfo{year}{2006}), \bibinfo{note}{cond-mat/0611342}.

\bibitem[{\citenamefont{McCann}(2006)}]{McC06}
\bibinfo{author}{\bibfnamefont{E.}~\bibnamefont{McCann}},
  \bibinfo{journal}{Phys. Rev. B} \textbf{\bibinfo{volume}{74}},
  \bibinfo{pages}{161403} (\bibinfo{year}{2006}).

\bibitem[{\citenamefont{Charlier et~al.}(1992)\citenamefont{Charlier,
  Michenaud, and Lambin}}]{CML92}
\bibinfo{author}{\bibfnamefont{J.~C.} \bibnamefont{Charlier}},
  \bibinfo{author}{\bibfnamefont{J.~P.} \bibnamefont{Michenaud}},
  \bibnamefont{and} \bibinfo{author}{\bibfnamefont{P.}~\bibnamefont{Lambin}},
  \bibinfo{journal}{Phys. Rev. B} \textbf{\bibinfo{volume}{46}},
  \bibinfo{pages}{4540} (\bibinfo{year}{1992}).

\bibitem[{\citenamefont{Pong and Durkan}(2005)}]{PD05_short}
\bibinfo{author}{\bibfnamefont{W.~T.} \bibnamefont{Pong}} \bibnamefont{and}
  \bibinfo{author}{\bibfnamefont{C.}~\bibnamefont{Durkan}},
  \bibinfo{journal}{J. of Phys. D} \textbf{\bibinfo{volume}{38}},
  \bibinfo{pages}{R329} (\bibinfo{year}{2005}).

\bibitem[{\citenamefont{Rong and Kuiper}(1993)}]{RK93}
\bibinfo{author}{\bibfnamefont{Z.~Y.} \bibnamefont{Rong}} \bibnamefont{and}
  \bibinfo{author}{\bibfnamefont{P.}~\bibnamefont{Kuiper}},
  \bibinfo{journal}{Phys. Rev. B} \textbf{\bibinfo{volume}{48}},
  \bibinfo{pages}{17427 } (\bibinfo{year}{1993}).

\bibitem[{\citenamefont{{Hass {\it et al.}}}(2007)}]{hass-2007_short}
\bibinfo{author}{\bibfnamefont{J.}~\bibnamefont{{Hass {\it et al.}}}}
  (\bibinfo{year}{2007}), \bibinfo{note}{cond-mat/0702540}.

\bibitem[{\citenamefont{{Hass {\it et al.}}}(2006)}]{HFL+06_short}
\bibinfo{author}{\bibfnamefont{J.}~\bibnamefont{{Hass {\it et al.}}}},
  \bibinfo{journal}{Appl. Phys. Lett.} \textbf{\bibinfo{volume}{89}},
  \bibinfo{pages}{143106} (\bibinfo{year}{2006}).

\bibitem[{\citenamefont{Guinea et~al.}(2006)\citenamefont{Guinea, {Castro
  Neto}, and Peres}}]{GNP06}
\bibinfo{author}{\bibfnamefont{F.}~\bibnamefont{Guinea}},
  \bibinfo{author}{\bibfnamefont{A.~H.} \bibnamefont{{Castro Neto}}},
  \bibnamefont{and} \bibinfo{author}{\bibfnamefont{N.~M.~R.}
  \bibnamefont{Peres}}, \bibinfo{journal}{Phys. Rev. B}
  \textbf{\bibinfo{volume}{73}}, \bibinfo{pages}{245426}
  (\bibinfo{year}{2006}).

\bibitem[{\citenamefont{{Bostwick \emph{et al.}}}(2006)}]{bostwick-2006_short}
\bibinfo{author}{\bibfnamefont{A.}~\bibnamefont{{Bostwick \emph{et al.}}}}
  (\bibinfo{year}{2006}), \bibinfo{note}{arXiv:cond-mat/0609660}.

\bibitem[{\citenamefont{{Zhou \emph{ et al.}}}(2006)}]{ZGG+06_short}
\bibinfo{author}{\bibfnamefont{S.~Y.} \bibnamefont{{Zhou \emph{ et al.}}}},
  \bibinfo{journal}{Nature Physics} \textbf{\bibinfo{volume}{2}},
  \bibinfo{pages}{595 } (\bibinfo{year}{2006}).

\bibitem[{\citenamefont{{Berger {\it et al.}}}(2006)}]{BSL+06_short}
\bibinfo{author}{\bibfnamefont{C.}~\bibnamefont{{Berger {\it et al.}}}},
  \bibinfo{journal}{Science} \textbf{\bibinfo{volume}{312}},
  \bibinfo{pages}{1191} (\bibinfo{year}{2006}).

\bibitem[{\citenamefont{{Sadowski {\it et al.}}}(2006)}]{SMP+06b_short}
\bibinfo{author}{\bibfnamefont{M.~L.} \bibnamefont{{Sadowski {\it et al.}}}},
  \bibinfo{journal}{Phys. Rev. Lett.} \textbf{\bibinfo{volume}{97}},
  \bibinfo{pages}{266405} (\bibinfo{year}{2006}).

\bibitem[{\citenamefont{de~{Heer {\it et al.}}}(2007)}]{deheer07}
\bibinfo{author}{\bibfnamefont{W.~A.} \bibnamefont{de~{Heer {\it et al.}}}}
  (\bibinfo{year}{2007}), \bibinfo{note}{cond-mat/0704.0285}.

\bibitem[{\citenamefont{{Tang {\it et al.}}}(1996)}]{PhysRevB.53.979-b_short}
\bibinfo{author}{\bibfnamefont{M.~S.} \bibnamefont{{Tang {\it et al.}}}},
  \bibinfo{journal}{Phys. Rev. B} \textbf{\bibinfo{volume}{53}},
  \bibinfo{pages}{979} (\bibinfo{year}{1996}).

\end{thebibliography}
%

\end{document}